\documentstyle[preprint,prb,aps]{revtex}
\begin{document}
\preprint{cond-mat/9502038}
\draft

\title{Experimental determination of the quasi-particle decay length
$\xi_{\text{Sm}}$\/ in a superconducting quantum well.\thanks{Submitted
to Phys. Rev. B, Rap. Comm.}}

\author{P.~H.~C.~Magn\'ee,  B.~J.~van Wees, and T.~M.~Klapwijk}

\address{Department of Applied Physics and Material Science Centre,
University of Groningen,\\ Nijenborgh 4, 9747 AG Groningen, The
Netherlands.}

\author{W. van de Graaf, and G. Borghs}

\address{Interuniversity Micro Electronics Centre, Kapeldreef 75,
B-3030, Leuven, Belgium}

\date{\today}

\maketitle

\begin{abstract}
We have investigated experimentally the electronic transport
properties of a two-dimensional electron gas (2DEG) present in an
AlSb/InAs/AlSb quantum well, where part of the toplayer has been
replaced by a superconducting Nb strip, with an energy gap
$\Delta_0$\/. By measuring the lateral electronic transport underneath
the superconductor, and comparing the experimental results with a
model based on the Bogoliubov-de Gennes equation and the
Landauer-B\"uttiker formalism, we obtain a decay length
$\xi_{\text{Sm}} \approx 100~\text{nm}$\/ for electrons. This decay
length corresponds to an interface transparency $T_{\text{SIN}}=0.7$\/
between the Nb and InAs. Using this value, we infer an energy gap in
the excitation spectrum of the SQW of $\Delta_{\text{eff}} = 0.97
\Delta_0 = 0.83~\text{meV}$\/.
\end{abstract}


\narrowtext

A superconducting quantum well (SQW) can be defined as a system in
which one of the barriers of a quantum well, in our case InAs in
between AlSb barriers, is replaced by a superconductor, here Nb. In a
quantum well, particles are confined by normal reflections at the
boundaries. In a SQW, also Andreev reflection \cite{And64} at the
superconducting barrier can occur, changing the confinement. Due to
this superconducting barrier, an energy gap $\Delta_{\text{eff}}$\/
appears in the excitation spectrum of the two dimensional electron gas
(2DEG) present in the SQW. \cite{VMvWK94} The magnitude of this gap
depends on the interface transparency $T_{\text{SIN}}$\/ between the
superconductor and the InAs, and the superconducting energy gap
$\Delta_0$\/. In the limit $T_{\text{SIN}}\ll 1$\/, Volkov {\em et
al.\/} \cite{VMvWK94} have shown that the SQW can be described as a 2
dimensional superconductor with an effective order parameter
$\Delta_{\text{eff}}\,e^{i\phi}$\/ ($\Delta_{\text{eff}} \ll
\Delta_0$\/), where $\phi$\/ is the macroscopic phase of the
superconductor on top.

The physics of the SQW is of importance to understand transport in
co-planar super-normal-superconductor (SNS) junctions. From a
technological point of view there are two kinds of SNS junctions,
sandwich- (inline planar) and co-planar junctions. \cite{Lik79} In
sandwich-type junctions, the junction length $L$\/ is well defined. In
co-planar structures however, electrons can travel a certain distance
underneath the superconductor before being Andreev reflected, thus
effectively enlarging the junction length. The distance electrons
penetrate underneath the superconductor can be associated with a decay
length $\xi_{\text{Sm}}$\/, similar to the superconducting coherence
length $\xi_0 = \hbar v_{\text{F}} / \Delta$\/.

It is important to have a good estimate of the actual junction length,
$L_{\text{eff}} = L + 2 \xi_{\text{Sm}}$, because it is a relevant
parameter in calculations for the critical current $I_c$\/ in SNS
junctions. \cite{Lik76,KL88} This was also appreciated recently
by Nguyen {\em et al.\/}, \cite{NKH92,NKH94} who measured Nb-InAs-Nb
junctions with varying length, using a transmission line model (TLM).
When plotting the resistance versus the junction length, they obtained
a straight line, intersecting the length axis at a negative value.
They interpreted this length to be the average distance
$x_{\text{A}} = 1.5~\mu$\/m an electron needs to travel underneath the
superconductor, before it is Andreev reflected.

The system under study is a 15~nm InAs layer sandwiched between a
2~$\mu$\/m AlSb layer and a superconductor, Nb, see
Fig~\ref{fig:sample}b. The 2DEG present in the InAs has a high
electron mobility, resulting in a long elastic mean free path
$\ell$\/. \cite{DHvWetal95} The ballistic regime is therefore easily
accessible. Furthermore the absence of a Schottky barrier in
metal-InAs contacts enables one to make highly transparent interfaces.
At the InAs-AlSb interface, the barrier for electrons is assumed to be
infinite, at the Nb-InAs interface a $\delta$\/-function potential
barrier is present, characterized by a dimensionless parameter $Z =
2p_{\text{F}}H/\mu = H/\hbar v_{\text{F}}$\/ as introduced by Blonder
{\em et al.\/} \cite{BTK82} Due to the high interface transparency in
our system the limit $\Delta_{\text{eff}} \ll \Delta_0$\/, used by
Volkov {\em et al.\/}, \cite{VMvWK94} no longer applies, therefore
also the quasi particle decay length is expected to be different from
$\hbar v_{\text{F}} / \Delta_{\text{eff}}$\/. The classical
description used by Nguyen {\em et al.\/} \cite{NKH92,NKH94} takes
into account multiple Andreev reflections, but ignores phase coherence
between these multiple reflections. This approach will in general not
lead to an exponential decay of the laterally transmitted wave
functions in the SQW. Here we will calculate both the energy gap and
the decay length of quasi particles at the Fermi-level in the SQW,
using a quantum mechanical description.

We will assume that only the lowest 2D subband in the QW is filled,
which is the case in our samples. In order to calculate the wave
functions we have to solve the Bogoliubov-de Gennes equation,
\cite{dGen66}
\begin{equation}
\left[
    \begin{array}{lr}
        {\cal H}        &\Delta({\bf r})\\
        \Delta^{\ast}({\bf r})&-{\cal H}
    \end{array}
\right]
\left[
    \begin{array}{c}
        u({\bf r})\\
        v({\bf r})
    \end{array}
\right]
=E \left[
       \begin{array}{c}
           u({\bf r})\\
           v({\bf r})
       \end{array}
   \right]\, ,
\label{eq:BdG}
\end{equation}
\noindent where the Hamiltonian $\cal H$\/ is defined as
\begin{equation}
{\cal H}=-\frac{\hbar^2}{2 m^{\ast}}\nabla^2+U({\bf r})-\mu\, .
\label{eq:ham}
\end{equation}
\noindent For the effective mass we assume $m^{\ast}=m_0$\/, the free
electron mass, in the Nb, and $m^{\ast}=0.023~m_0$\/ in the InAs. For
the potential $U({\bf r})$\/ we take
\begin{eqnarray}
U({\bf r}) = U(z) &=& H \delta(z) - E_{\text{F, Nb}} \theta(-z)
\nonumber\\
&-& E_{\text{F,InAs}} \theta(z)\theta(L-z) + V_0 \theta(z-L)\, ,
\label{eq:pot}
\end{eqnarray}
\noindent where the Fermi energies are $E_{\text{F, Nb}} =
5.3~\text{eV}$\/ and $E_{\text{F, InAs}} = 0.11~\text{eV}$\/. The
potential barrier at the InAs-AlSb interface, $V_0$\/, is assumed to
go to infinity. In Eq.(\ref{eq:BdG}) the pair potential $\Delta({\bf
r})$\/ is assumed to be $\Delta_0$ in the Nb $(z<0)$\/, and zero
elsewhere.

The solutions to the Bogoliubov-de Gennes equation, Eq.(\ref{eq:BdG}),
are given by electron- and hole wave functions in the InAs-quantum
well $(0<z<L)$\/,
$$
\Psi({\bf r})=
\left[
    \begin{array}{c}
        u({\bf r})\\
        v({\bf r})
    \end{array}
\right]
=
\left\{
u(z)
  \left[
      \begin{array}{c}
          1\\
          0
      \end{array}
  \right]
+ v(z)
  \left[
      \begin{array}{c}
          0\\
          1
      \end{array}
  \right]
\right\}
e^{i(k_xx+k_yy)}\, ,
$$
and by mixed quasi-particle wave functions in the Nb superconductor
$(z<0)$\/, with $u^2 = 1 - v^2 = \frac{1}{2} (1+\Omega/E)$\/.
$\Omega^2 = E^2 - \Delta^2$\/, $E$\/ being the energy with respect to
the Fermi-level. These wave functions and their derivatives have to be
matched at the boundaries $z=0$\/ and $z=L$\/, see
Ref.~\onlinecite{VMvWK94} for a detailed analysis. Numerical solutions
of Eq.(\ref{eq:BdG}) are given in Fig.~\ref{fig:matrix}. For $k_z$\/
real, where $k_z$\/ is the wave vector in the $z$\/-direction, there
are no solutions of $E(k)$\/ with $|E|$\/ smaller than an effective
energy gap $|\Delta_{\text{eff}}|$\/. This $|\Delta_{\text{eff}}|$\/
is calculated as function of the transparency $T_{\text{SIN}} =
1/(1+Z^2)$\/ of the Nb-InAs interface, by finding the minimum of
$E(k)$\/, Fig.~\ref{fig:matrix}a. At low transparency it can be shown
\cite{VMvWK94} that $\Delta_{\text{eff}}$\/ depends linearly on
$T_{\text{SIN}}$\/: $\Delta_{\text{eff}} \approx \frac{1}{4}
T_{\text{SIN}} E_0$\/, where $E_0 = \frac{\hbar^2}{2 m^{\ast}}
\left(\frac{\pi}{L}\right)^2$\/ is the confinement energy in the QW.
At $E=0$\/ there are only solutions for Eq.(\ref{eq:BdG}) for complex
$k_z$\/. The total energy $\frac{\hbar^2}{2 m^{\ast}} (k_z^2 +
k_{\parallel}^2)$\/ must be real, hence $k_{\parallel} = \sqrt{k_x^2 +
k_y^2}$\/ has an imaginary part. By taking the $y$\/-direction along
the boundary of the SQW, wave function matching requires
$\text{Im}(k_y) = 0$\/. For the decay length we can thus write
$\text{Im}(k_x)^{-1} = -\text{Re}(k_x) / \text{Re}(k_z)\text{Im}(k_z)
\approx -k_{\text{F}} \cos(\alpha) / \text{Re}(k_z)\text{Im}(k_z) =
\xi_{\text{Sm}} \cos(\alpha)$\/, where $\alpha$\/ is the angle of
incidence. The decay length $\xi_{\text{Sm}}$\/ is plotted in
Fig.~\ref{fig:matrix}b. The decay length $\xi = \hbar^2 k_{\text{F}} /
m^{\ast} \Delta_{\text{eff}}$\/, analogous to the expression used for
the decay length of quasi-particles in a superconductor, is shown for
comparison. As can be seen, at high transparency, there is a
substantial difference between both decay lengths. We will show that
the former one is the relevant one in the SQW.

Samples are based on a 15~nm InAs quantum well with AlSb barriers.
Prior to any processing the top AlSb layer is removed. The parameters
of the quantum well with an exposed InAs surface are (measured at
4.2~K): $n_{\text{S}} = 1.1 \times 10^{16}~\text{m}^{-2}$\/, $\mu_e =
2.2~\text{m}^2/\text{Vs}$\/ and $\ell = 380~\text{nm}$\/. We want to
measure the laterally transmitted signal through a SQW, depending on
the width $d$\/. For this purpose, the Nb pattern is defined, using
electron beam lithography (EBL), as a narrow strip, either $d = 100$\/
or 200~nm, with probes at either side at distances of 200 and 300~nm,
see Fig.~\ref{fig:sample}a. Inspection with the electron microscope
shows $d = 100$\/ and 236~nm. Prior to the Nb deposition, the InAs
surface is cleaned using low energy Ar-sputtering. This can reduce the
thickness of the quantum well by a maximum of 2~nm, and might alter
the carrier density $n_{\text{S}}$\/, and mobility $\mu_e$\/. To
define the width of the InAs channel, a mesa-etch is performed in the
200~nm strip sample, $W = 0.9~\mu$\/m. This was not done for the
sample with a 100~nm strip, which means that in this sample the
junctions to the SQW have a width equal to the length of the strip,
3.5~$\mu$\/m. Evidently some parallel conductance will be present.

Prior to the measurements we checked the continuity of the strips,
together with the critical temperature $T_c$\/, by measuring the
resistance through contacts 3 and 4 (Fig.~\ref{fig:sample}a).
Measurements are done using a standard 4-point lock-in technique, at
1.3~K. By applying an ac modulation current on top of a dc-bias, we
can measure the energy dependence of transport in the SQW.
\cite{footnote1} Transport through the SQW can be modelled in the
spirit of the Landauer-B\"uttiker formalism, \cite{BILP85,Lam91} using
normal- and Andreev reflection probabilities, $R_{e\rightarrow e}$\/
and $R_{e\rightarrow h}$\/, and transmission probabilities
$T_{e\rightarrow e}$\/ and $T_{e\rightarrow h}$\/, see
Fig~\ref{fig:sample}b. Conservation of particles requires $1 =
R_{e\rightarrow e} + R_{e\rightarrow h} + T_{e\rightarrow e} +
T_{e\rightarrow h}$\/. By expressing the current in terms of these
reflection and transmission probabilities, we can translate the region
underneath the strip into a schematic resistor network, shown in
Fig.~\ref{fig:sample}c, where
\begin{mathletters}
\begin{equation}
R = \frac{1}{G_{\text{S}}} \frac{1}{2 (R_{e\rightarrow h} +
        T_{e\rightarrow e})}\, ,
\end{equation}
\begin{equation}
R_C = \frac{1}{G_{\text{S}}} \frac{(T_{e\rightarrow e} -
        T_{e\rightarrow h})}{4 (R_{e\rightarrow h} + T_{e\rightarrow
        h}) (R_{e\rightarrow h} + T_{e\rightarrow e})}\, .
\end{equation}
\end{mathletters}
\noindent Here $G_{\text{S}} = \frac{2e^2}{h} \frac{W}{\frac{1}{2}
\lambda_{\text{F}}}$\/ is the Sharvin conductance. \cite{BvH91a}
Although these resistors do not have physical relevance, they are
useful for calculating the the various measurable quantities:
\begin{mathletters}
\begin{eqnarray}
&&\frac{\partial V_1}{\partial I_1} \approx \frac{\partial
V_2}{\partial I_2} \approx \frac{R_{\parallel}}{R_{\parallel} + R +
R_C} (R + R_C) = \frac{R_{\parallel}}{R_{\parallel} + R +
R_C}\nonumber\\
&&\times \frac{1}{G_{\text{S}}} \frac{2 (R_{e\rightarrow h} +
T_{e\rightarrow h}) + (T_{e\rightarrow e} - T_{e\rightarrow h})}{4
(R_{e\rightarrow h} + T_{e\rightarrow h}) (R_{e\rightarrow h} +
T_{e\rightarrow e})} \label{eq:dv1di1}\, ,
\end{eqnarray}
\begin{eqnarray}
&&\frac{\partial V_2 / \partial I_1}{\partial V_1 / \partial I_1}
\approx \frac{R_{\parallel}}{R_{\parallel} + R + R_C} \frac{R_C}{R +
R_C}= \frac{R_{\parallel}}{R_{\parallel} + R + R_C}\nonumber\\
&&\times \frac{(T_{e\rightarrow e} - T_{e\rightarrow h})}{2
(R_{e\rightarrow h} + T_{e\rightarrow h}) + (T_{e\rightarrow e} -
T_{e\rightarrow h})}\label{eq:dv2dv1}\, ,
\end{eqnarray}
\end{mathletters}
\noindent where $R_{\parallel}$\/ accounts for any parallel
conductance that might be present. For small $(T_{e\rightarrow e} -
T_{e\rightarrow h})$\/, the angular distribution of incoming electrons
is taken into account by calculating the following integral:
\begin{eqnarray}
T_{e\rightarrow e} &-& T_{e\rightarrow h} =
\frac{|\Psi(x=d)|^2}{|\Psi(x=0)|^2} \nonumber\\
&=& \frac{1}{2} \int_{-\pi/2}^{\pi/2} \cos(\alpha)
\exp\left(\frac{-2d}{\xi_{\text{Sm}}\cos(\alpha)}\right)\, d\alpha\, .
\label{eq:Tee}
\end{eqnarray}
\noindent In Fig.~\ref{fig:exp} the experimental data are shown. For
the wide strip sample, $d = 236~\text{nm}$\/, we measure junction
resistances of $\partial V_1 / \partial I_1 |_{V_1=0} = 425~\Omega$\/
and $\partial V_2 / \partial I_2 |_{V_2=0} = 625~\Omega$\/, whereas
from the Sharvin conductance, with $W = 0.9~\mu$\/m, we would expect
$1/G_{\text{S}} \simeq 180~\Omega$\/, which is approximately a factor
of 3 smaller. This can be explained by elastic scattering, present in
the samples. In the limit $T_{e\rightarrow e},\, T_{e\rightarrow h}
\ll 1$\/, which we assume to be the case, we can write $\partial
V_1/\partial I_1 \approx \frac{1}{G_{\text{S}}} \frac{1}{2
R_{e\rightarrow h}}$\/, Eq.(\ref{eq:dv1di1}). Using $\partial V_{1(2)}
/ \partial I_{1(2)} \approx 525~\Omega$\/ we get $R_{e\rightarrow h}
\approx 0.17$\/, or $R_{e\rightarrow e} \approx (1 - R_{e\rightarrow
h}) \approx 0.83$\/. This value for $R_{e\rightarrow h}$\/ is used in
further analysis. For the narrow strip sample, $d=100~\text{nm}$\/,
the junction width is $W=3.5~\mu$\/m. By scaling $\partial V_{1(2)} /
\partial I_{1(2)}$\/ from the wide strip sample, we expect $\partial
V_{1(2)} / \partial I_{1(2)} \approx 135~\Omega$\/. The measured
values are $\partial V_1 / \partial I_1 |_{V_1=0} = 75~\Omega$\/ and
$\partial V_2 / \partial I_2 |_{V_2=0} = 68~\Omega$\/. These lower
values are explained by a parallel resistance of $R_{\parallel}
\approx 150~\Omega$, which is in agreement with expectation on
geometrical grounds. We will first focus on the zero voltage bias
transfer signal $\partial V_2 / \partial V_1$\/. From the wide strip,
$d = 236~\text{nm}$\/, we obtain $\partial V_2 / \partial V_1|_{V_1=0}
= \partial V_1 / \partial V_2|_{V_2=0} \approx 0.01$\/ which, with the
aid of Eq.(\ref{eq:dv2dv1}) leads to $(T_{e\rightarrow e} -
T_{e\rightarrow h}) \approx 0.0035$\/. From Eq.(\ref{eq:Tee}) we can
then calculate $\xi_{\text{Sm}} \approx 100~\text{nm}$\/. Using again
Eq.(\ref{eq:Tee}) and (\ref{eq:dv2dv1}) for the narrow strip, $d =
100~\text{nm}$\/, we expect $\partial V_2 / \partial V_1 \approx
0.097$\/, which is in reasonable agreement with the observed values
$\partial V_2 / \partial V_1|_{V_1=0} \approx 0.04$\/ and $\partial
V_1 / \partial V_2|_{V_2=0} \approx 0.05$\/. The discrepancy is mainly
due to the fact that we assume $R_{e\rightarrow h}$\/ to be equal for
both samples, whereas in general it will depend on $d$\/. Using
$\xi_{\text{Sm}} \approx 100~\text{nm}$\/ we can estimate from
Fig.~\ref{fig:matrix}b that the transparency $T_{\text{SIN}} = 0.7$\/
for the Nb-InAs interface.

At finite energy, $\xi_{\text{Sm}} \propto \text{Im}(k_x)^{-1}$\/ will
increase with $E$\/, until $E \geq \Delta_{\text{eff}}$\/, where the
decay length will diverge, for $E>\Delta_{\text{eff}}$\/ there will be
propagating states. Therefore the transfer signal $\partial V_2 /
\partial V_1 (V_1)$\/ is expected to increase with increasing bias
voltage $V_1$\/. In both samples we observe a dip at low voltage bias,
with a width of approximately 0.8~mV. From the estimated interface
transparency $T_{\text{SIN}} = 0.7$\/ we infer an induced energy gap
$\Delta_{\text{eff}} = 0.97 \Delta_0 = 0.83~\text{meV}$\/, where
$\Delta_0 = 0.86~\text{meV}$\/ is the superconducting energy gap of
the strip, which is very close to width of the observed dip. This dip
is however only observed when the current is injected from one probe,
when the current is injected from the opposite probe, there is no dip
present in the transfer signal, see Fig.~\ref{fig:exp}. This indicates
that we can not understand the voltage dependence within the presented
model. When the elastic scattering length is small, $\ell \leq
\xi_{\text{Sm}}$\/, we do not measure $\xi_{\text{Sm}}$\/, but
$\xi_{\text{eff}} = \sqrt{\xi_{\text{Sm}} \ell}$\/. This will lead to
a somewhat larger value for $\xi_{\text{Sm}}$\/. In the InAs used
$\ell$\/ is reduced due to the Ar-sputtering, and we expect that this
will influence especially the voltage dependent signal. Furthermore we
see a small dip at $V = 2~\text{mV} \approx (\Delta_{0,probe} +
\Delta_{0,strip}) / e$\/. This is not expected from our model, but is
probably related to the fact that we do not inject electrons from a
normal reservoir, but use a superconducting injector instead.
\cite{footnote1}

Nguyen {\em et al.\/} \cite{NKH94} obtained an Andreev transfer length
$x_{\text{A}} = 1.5~\mu$\/m, which is equivalent to a decay length for
the wave functions of 3~$\mu$\/m. Comparing this value to the decay
length obtained from our experiment, $\xi_{\text{Sm}} \approx
100~\text{nm}$\/, we see a huge difference. Nguyen {\em et al.\/}
obtain their $x_{\text{A}}$\/ by extrapolating resistances of long
Nb-InAs-Nb junctions (20 to 200~$\mu$\/m) to lower junction lengths.
This is however not allowed, since a junction of zero length still has
a finite resistance, of the order of the Sharvin resistance.
\cite{BvH91a}

In conclusion, we have investigated lateral transport in a
Nb-InAs-AlSb superconducting quantum well. At zero voltage bias we can
describe the transport properties in terms of transmission and
reflection probabilities. For quasi-particles in the SQW, a decay
length of $\xi_{\text{Sm}} \approx 100~\text{nm}$\/ is inferred from
the experiments. This $\xi_{\text{Sm}}$\/ corresponds to an interface
transparency between the Nb and InAs of $T_{\text{SIN}}=0.7$\/, which,
according to calculations, results in an induced energy gap in the
excitation spectrum of the Nb-InAs SQW of $\Delta_{\text{eff}} = 0.97
\Delta_0$\/. From the experiment there are some indications for the
presence of this gap, although the exact voltage dependence of the
transfer signal can not be understood within the presented model. We
believe that the transfer signal at finite bias is greatly influenced
by elastic scattering.

This work was supported by the Dutch Science Foundation NWO/FOM. B.~J.
van Wees acknowledges support from the Royal Dutch academy of Sciences
(KNAW).


\begin{thebibliography}{10}
\bibitem{footnote1}
Due to the highly non-uniform density of states in a superconductor,
the energy distribution of electrons injected from a superconductor
will in general be non-linear in voltage bias.
\end{thebibliography}

\begin{figure}
\caption{The normalized induced energy gap
$\Delta_{\text{eff}}/\Delta_0$\/ (Eq.(\protect\ref{eq:BdG}), $\min
E(k)$\/) as function of the barrier tranmittance
$T_{\text{SIN}}=1/(1+Z^2)$\/ of the Nb-InAs interface (a), the dashed
line shows the linear dependence for $T_{\text{SIN}} \ll 1$\/.
\protect\cite{VMvWK94} Figure (b) shows the calculated decay length,
$\xi_{\text{Sm}}=|k_{\text{F}} / (\text{Re}(k_z) \text{Im}(k_z))|$\/
(Eq.(\protect\ref{eq:BdG}), $E=0$\/). The quasi-particle decay
length $\xi = \hbar k_{\text{F}} / m^{\ast} \Delta_{\text{eff}}$\/
analogous to that of a superconductor is given for comparison (dashed
curve). For these curves we used $k_{\text{F}}=0.26~\text{nm}^{-1}$\/,
and $\Delta_0 = 1.0~\text{meV}$\/ ($T_c \sim 6~\text{K}$\/).}
\label{fig:matrix}
\end{figure}

\begin{figure}
\caption{Schematic picture of the sample under study, a narrow strip,
width $d$\/, with two probes on either side (a). The distances of the
probes to the strip are 200 (1) and 300~nm (2). The width of the InAs
channel is defined by a mesa-etch. In the absence of a mesa-etch, the
length of the strip determines the junction width. For electrons
underneath the narrow strip we can define normal- and Andreev-like
reflection and transmission coefficients $R_{e\rightarrow
e,(e\rightarrow h)}$\/ and $T_{e\rightarrow e,(e\rightarrow h)}$\/
(b). This region can be represented by a simple network of resistors
$R$\/ and $R_C$\/ (c). In the absence of a mesa-etch, some parallel
conductance, characterized by $R_{\parallel}$\/, is present.}
\label{fig:sample}
\end{figure}

\begin{figure}
\caption{Differential resistance, $\partial V_1 / \partial I_1$\/ and
$\partial V_2 / \partial I_2$\/, of the junctions formed by probe 1
and the strip, and probe 2 and the strip, as function of the bias, and
the transmitted signals $\frac{\partial V_2 / \partial I_1}{\partial
V_1 / \partial I_1}$\/ and $\frac{\partial V_1 / \partial
I_2}{\partial V_2 / \partial I_2}$\/, for a strip with $d =
100~\text{nm}$\/ and a strip with $d = 236~\text{nm}$\/. All curves
are measured at $T=1.3~\text{K}$\/.}
\label{fig:exp}
\end{figure}

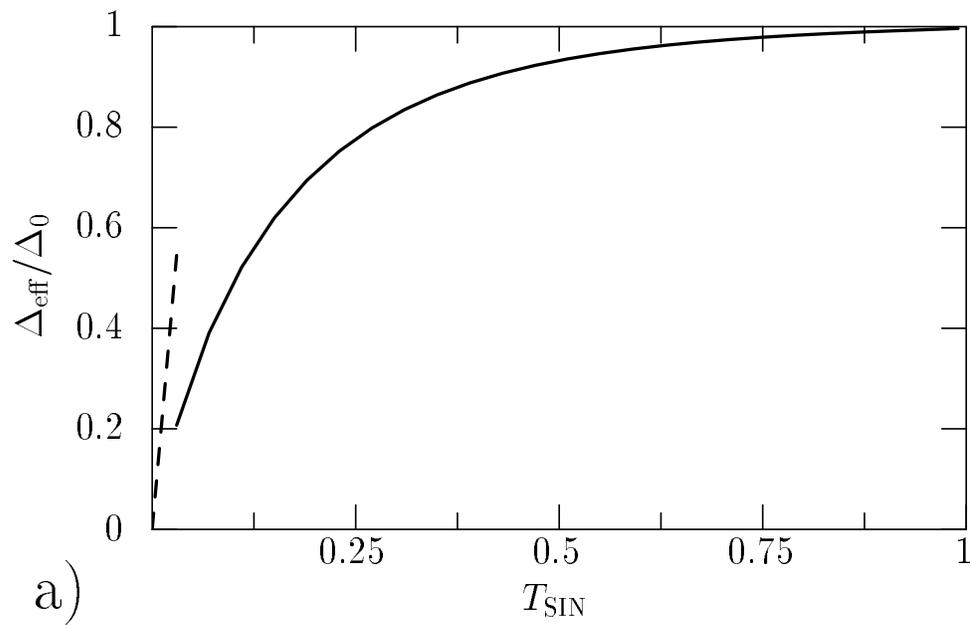
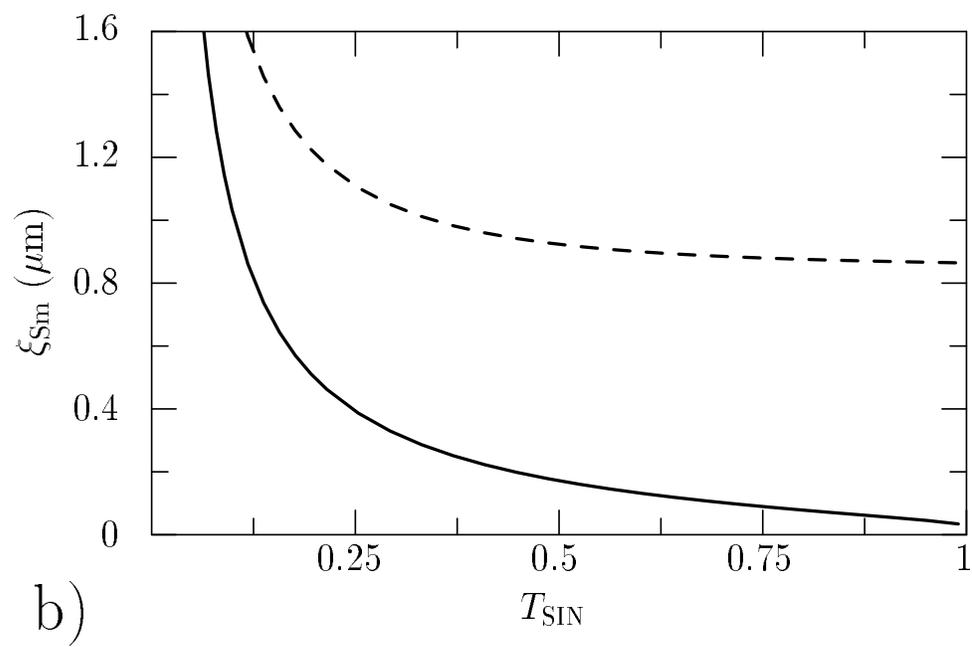
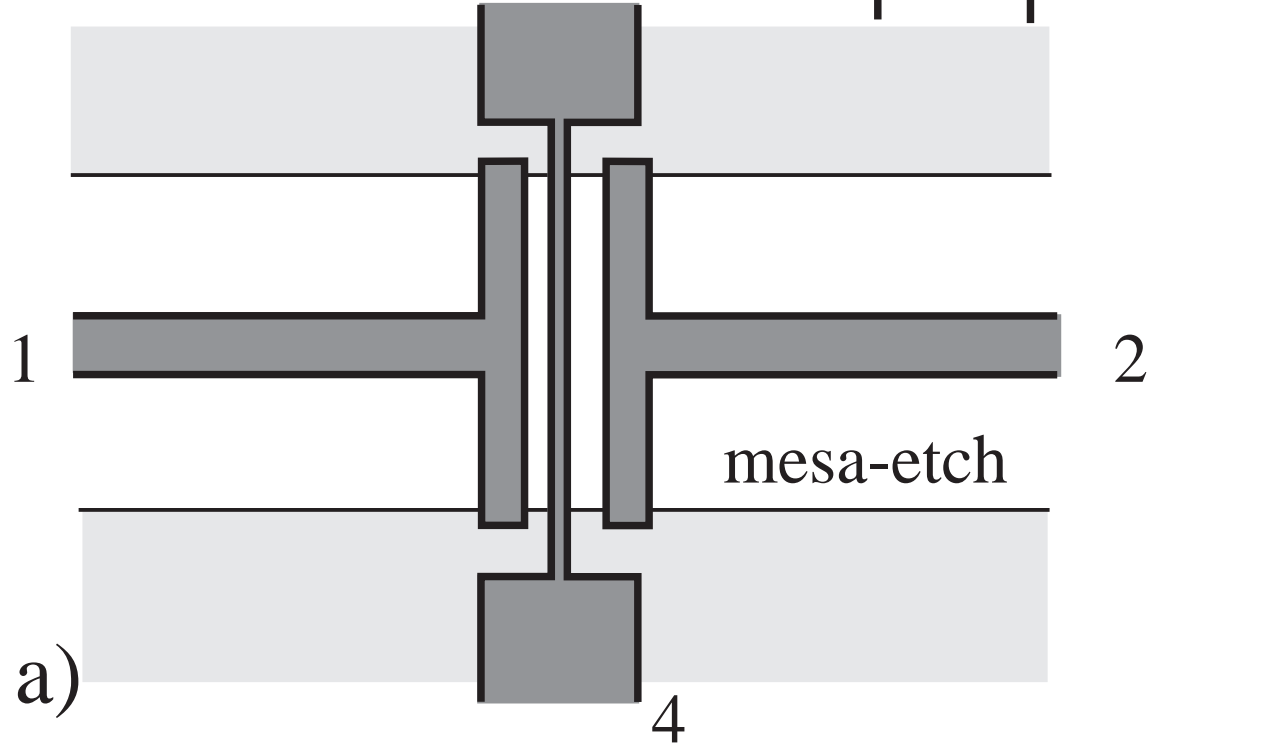
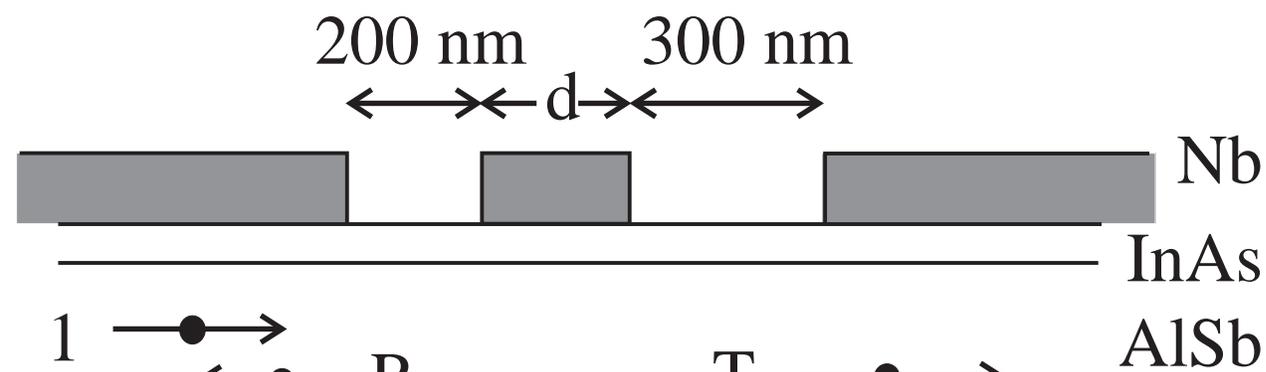
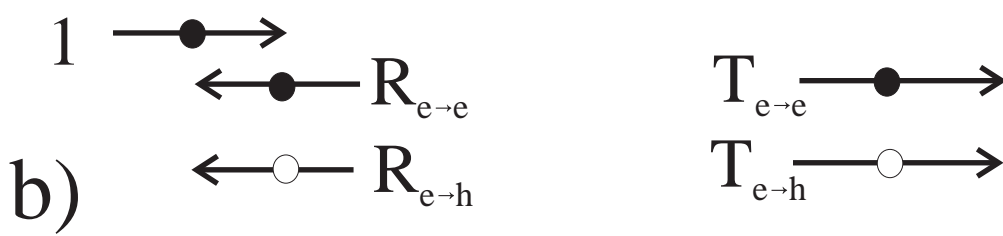
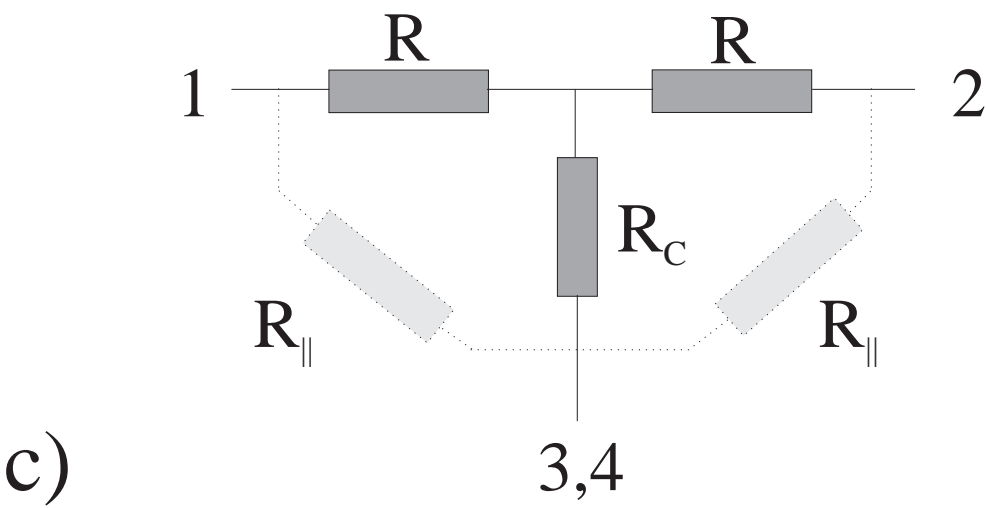
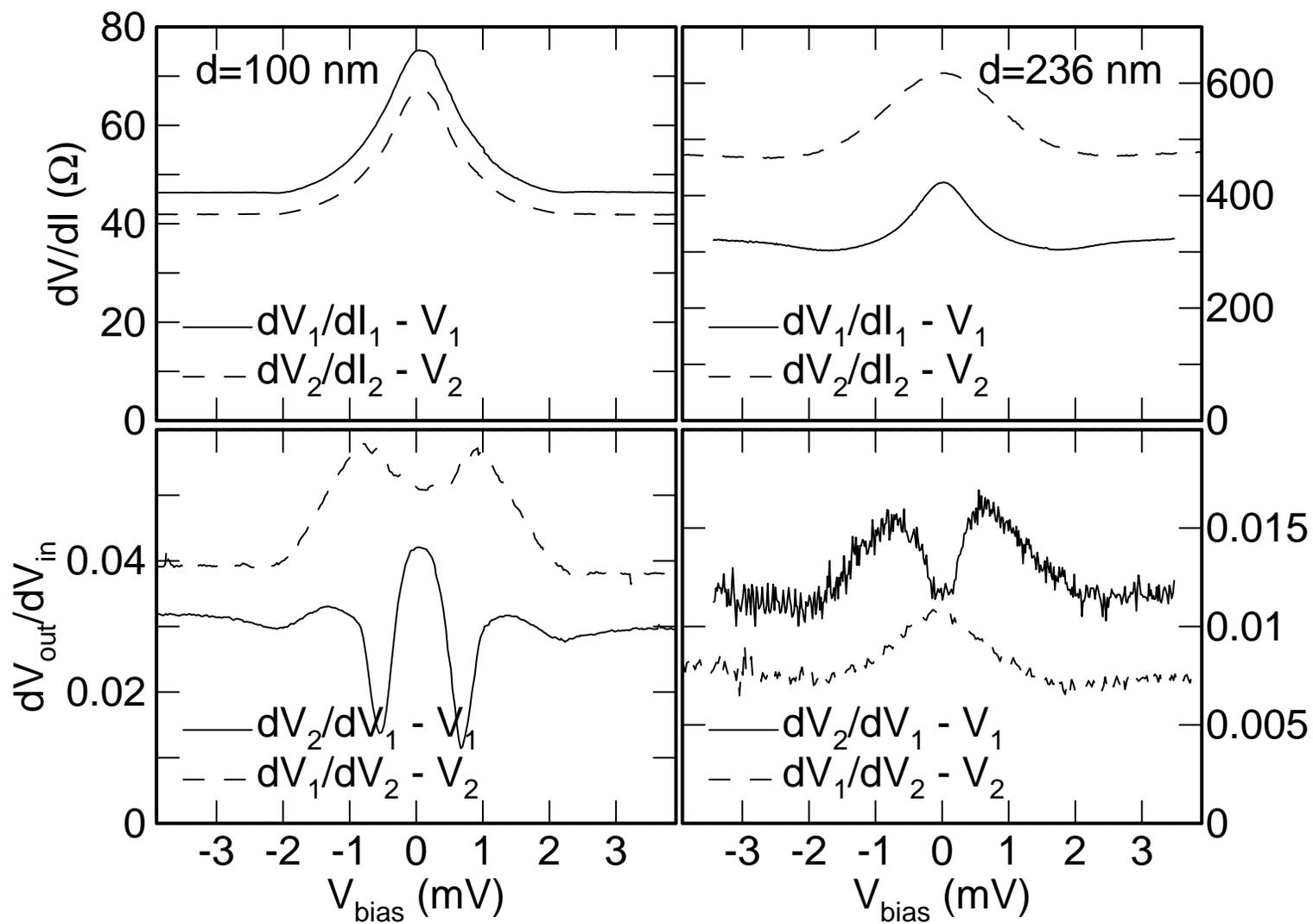
\end{document}